\documentstyle[12pt]{article}
\newcommand{\be}{\begin{equation}}
\newcommand{\ee}{\end{equation}}
\newcommand{\al}{\alpha}
\newcommand{\bt}{\beta}

\newcommand{\bi}{\bibitem}
\newcommand{\ve}{\varepsilon}

\begin{document}
\begin{titlepage}
\null

\begin{flushright}
KOBE-TH-94-04\\
hep-th/9409152\\
September 1994
\end{flushright}
\vspace{7mm}

 \begin{center}
  {\Large\bf Analysis of the Wheeler-DeWitt Equation\par}
  {\Large\bf beyond Planck Scale and Dimensional Reduction\footnote{
    Work supported by the Grant-in-Aid for Scientific Research
    from the Ministry of Education, Science and Culture (No.30183817).
    }\par}
 \end{center}
  \vspace{1.0cm}
  \baselineskip=7mm

\begin{center}
  {\large Tsutomu Horiguchi\footnote{
     E-mail address: horiguchi@phys02.phys.kobe-u.ac.jp} \par}
  \vspace{3mm}
   {\sl Graduate School of Science and Technology, Kobe University\\
        Rokkodai, Nada, Kobe 657, Japan \par}
  \vspace{8mm}
  {\large Kayoko Maeda\footnote{
     E-mail address: maeda@phys02.phys.kobe-u.ac.jp}
   and Makoto Sakamoto\footnote{
     E-mail address: sakamoto@phys02.phys.kobe-u.ac.jp} \par}
  \vspace{3mm}
   {\sl Department of Physics, Kobe University\\
        Rokkodai, Nada, Kobe 657, Japan\par}

\vspace{1.5cm}
{\Large\bf Abstract}
\end{center}
\par

We solve the Wheeler-DeWitt equation for {\it four}-dimensional
Einstein gravity
as an expansion in powers of the Planck mass
by means of a heat kernel regularization.
Our results suggest that
in the universe with a very small radius or with a very large curvature
beyond a Planck scale
expectation values of operators are
reduced to calculations in a path integral representation of
{\it three}-dimensional Einstein gravity.

\end{titlepage}
\setcounter{footnote}{0}
\baselineskip=7mm


As the evolution of the universe is followed backwards in time,
we would expect quantum gravitational effects to become important.
Nonperturbative effects would drastically change the notion of space-time
beyond the Planck scale and invalidate perturbative calculations
around a fixed background metric.
Quantum geometrodynamics based on the Wheeler-DeWitt equation \cite{W-DW}
is a natural framework which may answer the question of
quantum space-time structure at the early stage of the universe.
A lot of work has been done along this line and
various interesting results have been obtained
\cite{W-DW}-\cite{Horiguchi}.
However, in contrast with its grand motivation,
most of studies have been devoted on simple minisuperspace models
in the semiclassical approximation.
Short distance physics beyond the Planck
scale is not still uncovered.\footnote{
A new approach to canonical quantum gravity
has been proposed by Ashtekar \cite{Ashtekar}.
In terms of Ashtekar's new variables,
a large class of solutions to the Hamiltonian constraint
has been constructed in the loop representation \cite{loop},
though physical meanings of those solutions are still unclear.}
In this paper,
we propose a new approximation scheme to solve
the Wheeler-DeWitt equation and discuss physical meanings of
the wave function of the universe to leading order.
Our results may suggest that beyond the Planck scale dimensional
reduction of four-dimensional Einstein gravity occurs
and a new phase appears.

The (unregulated) Wheeler-DeWitt equation without matter is
\be
\biggl[\, -\frac{16 \pi}{m_p^2}\, G_{ijkl}(x)
    \frac{\delta}{\delta h_{kl}(x)}
    \frac{\delta}{\delta h_{ij}(x)}
+ \frac{m_p^2}{16 \pi}\,\sqrt{h(x)}
     \Bigl(\!~^{3}\!R(x) + 2 \Lambda \Bigr) \biggr]\,
      \Psi[h] = 0\   ,
\label{1}
\ee
where $m_p$ is the Planck mass and
$G_{ijkl}$ is the metric on superspace
\be
G_{ijkl}= \frac{1}{2 \sqrt{h}} (h_{ik} h_{jl} + h_{il} h_{jk}
            - h_{ij} h_{kl})\ .
\label{2}
\ee
The $~^{3}\!R(x)$ denotes the scalar curvature constructed from
the three-metric $h_{ij}$.
The Wheeler-DeWitt equation stands in need of regularization
because it contains a
product of two functional derivatives at the same spatial point,
\be
\Delta(x) \equiv G_{ijkl} (x)
  \frac{\delta}{\delta h_{kl}(x)}
  \frac{\delta}{\delta h_{ij}(x)}\ .
\label{3}
\ee
For example, $\Delta(x)$ acting on $~^{3}\!R(y)$ is proportional to
$(\delta(x,y))^2$, which is meaningless.
To make eq.(\ref{1}) well defined, we want to replace $\Delta(x)$ by
a renormalized operator $\Delta_{R}(x)$, which is a finite operator
preserving the three-dimensional general coordinate invariance.
Recently Mansfield proposed a renormalization scheme
to solve the Schr\"odinger equation for
Yang-Mills theory in the strong coupling
expansion \cite{Mans}.
We shall generalize
the renormalization procedure developed  by Mansfield
to the Wheeler-DeWitt equation and solve it for the wave function
of the universe as an expansion in powers of the Planck mass.

The first step to construct $\Delta_{R}(x)$ is to
\lq\lq point split"
the functional derivatives\cite{Mans}.
Consider the differential operator
\be
\Delta(x;t) \equiv
   \int d^3x' K_{i'j'kl}(x',x;t)
      \frac{\delta}{\delta h_{kl}(x)}
      \frac{\delta}{\delta h_{i'j'}(x')}\ ,
\label{4}
\ee
where $K_{i'j'kl}(x',x;t)$ is a bi-tensor
at both $x'$ and $x$ and satisfies the heat equation,
\be
- \frac{\partial}{\partial t}  K_{i'j'kl}(x',x;t) =
-{\nabla'}\!\!_{p} {\nabla'}^{p}  K_{i'j'kl}(x',x;t)\ ,
\label{5}
\ee
with the initial condition
\be
 \lim_{t \rightarrow 0}  K_{i'j'kl}(x',x;t) = G_{i'j'kl}(x) \delta(x',x)\ .
\label{6}
\ee
Here, ${\nabla'}\!\!_{p}$ and $\delta(x',x)$ denote the covariant
derivative with respect to $x'$ and the three-dimensional $\delta$ function,
respectively. The heat equation (\ref{5}) may be solved by the standard
technique\cite{S-DW}. Taking $t$ small but nonzero in eq.(\ref{4}) gives a
regulated operator  of $\Delta(x)$.
Let ${\cal O}$ be three-dimensional integrals of local functions of $h_{ij}$.
The action of $\Delta(x;t)$ on ${\cal O}$ may contain inverse powers of $t$
, which diverge as $ t \rightarrow 0$.
These powers of $t$ may be determined from dimensional analysis and
general coordinate invariance. We have, for example
\begin{eqnarray}
\Delta(x;t) \int d^3y \sqrt{h(y)} &=&
         \sqrt{h(x)} \biggl\{  \frac{\al_1}{t^{3/2}}
              + \frac{\al_2}{t^{1/2}}~^{3}\!R(x)
              + O(t^{1/2}) \biggr\}\  ,\nonumber\\
\Delta(x;t) \int d^3y \sqrt{h(y)}~^{3}\!R(y)
   &=&  \sqrt{h(x)} \biggl\{  \frac{\bt_1}{t^{5/2}}
           + \frac{\bt_2}{t^{3/2}} ~^{3}\!R(x) \nonumber\\
+ \frac{1}{t^{1/2}} \Bigl(  \bt_3 (\!~^{3}\!R(x))^2
  + \!\!\!\!&\bt_4&\!\!\!\!\!\!\! ~^{3}\!R_{ij}(x)\! ~^{3}\!R^{ij}(x)
           + \bt_5 \nabla\!_i \nabla^i\! ~^{3}\!R(x)  \Bigr)
           + O(t^{1/2}) \biggr\},
\label{7}
\end{eqnarray}
where $\al_n$'s and $\bt_n$'s are numerical constants.
The first few coefficients are given by
\begin{eqnarray}
   \al_1 &=& - \frac{21}{ 8 (4 \pi)^{3/2}}\ ,\nonumber\\
   \bt_1 &=&   \frac{ 3}{ 2 (4 \pi)^{3/2}}\ ,\nonumber\\
   \bt_2 &=& - \frac{11}{24 (4 \pi)^{3/2}}\ .
\label{8}
\end{eqnarray}
We cannot simply replace $\Delta(x)$ by $\lim_{t \rightarrow 0} \Delta(x;t)$
because of divergences of inverse powers of $t$.

The second step is to extract
a finite part from $ \Delta(x;t=0) {\cal O}$.
We again follow  the procedure by Mansfield.
We introduce a differentiable function $\phi(\varepsilon)$ such that
$\phi(\varepsilon)$
rapidly decreases to zero at infinity with $\phi(0) = 1 $.
Then, our definition of $\Delta_{R}(x)$ is
\be
\Delta_{R}(x) {\cal O} \equiv \lim_{s \rightarrow 0} F(s)\ ,
\label{9}
\ee
where
\be
F(s) = s \int^{\infty}_{0} d\ve\, \ve^{s-1} \phi(\ve)
         \Delta(x;t = \ve^2)\, {\cal O}\ .
\label{10}
\ee
We can see that the right hand side of eq.(\ref{9})
is equal to $\Delta(x;\ve^2=0) {\cal O}$
if we naively take the limit $s \rightarrow 0$.
By analytic continuation,
we can give a meaning to the right hand side of eq.(\ref{9})
even if $\Delta(x;\ve^2) {\cal O}$ diverges at the origin like $\ve^{-n}$
with integer $n$, and use the right hand side of eq.(\ref{9})
as our definition of $\Delta(x;\ve^2=0) {\cal O}$.
The integral $F(s)$ exists for $s > n$
( provided $\phi(\ve) \Delta(x; \ve^2) {\cal O}$ has no other divergences )
so that we can analytically continue $F(s)$ to small values of $s$
and take the limit $s \rightarrow 0$ to obtain a finite result.
For example, we have
\begin{eqnarray}
  \Delta_{R}(x) \int d^3y \sqrt{h(y)} &=&
      \sqrt{h(x)} \biggl\{  \al_1 \frac{ \phi^{(3)}(0) }{ 3! }
              + \al_2 \phi^{(1)}(0) ~^{3}\!R(x)  \biggr\}\  ,
               \nonumber\\
  \Delta_{R}(x) \int d^3y \sqrt{h(y)} ~^{3}\!R(y) &=&
     \sqrt{h(x)} \biggl\{  \bt_1 \frac{\phi^{(5)}(0)}{5!}
              + \bt_2 \frac{\phi^{(3)}(0)}{3!} ~^{3}\!R(x)\nonumber\\
  + \phi^{(1)}(0) \Bigl(  \bt_3 (\!~^{3}\!R(x))^2
     \!\! &+&\!\! \bt_4\! ~^{3}\!R_{ij}(x)\! ~^{3}\!R^{ij}(x)
                               + \bt_5 \nabla\!_i \nabla^i\! ~^{3}\!R(x)
            \Bigr)\biggr\}\  ,
\label{11}
\end{eqnarray}
where $\phi^{(n)}(0) \equiv \frac{d^n \phi(0)}{d \ve^n}$.
The results depend on the arbitrary function $\phi$.
This is an inevitable consequence of isolating finite quantities
from divergent ones.
Physical quantities must be independent of this arbitrariness,
so that coupling \lq\lq constants"
should be regarded as functionals of $\phi$.
This is the basic problem of renormalization.
We shall return to this point later.

As discussed above, we have the finite version of the Wheeler-DeWitt equation,
\be
   \biggl[\,  - \frac{16\pi}{m_p^2}\, \Delta_{R}(x)
   +\frac{m_p^2}{16\pi}\sqrt{h(x)}
     \Bigl(\!~^{3}\!R(x) + 2 \Lambda \Bigr) \biggr]\, \Psi[h] = 0\  .
\label{12}
\ee
We have chosen the renormalization procedure to preserve three-dimensional
general coordinate invariance but this is not enough to preserve the whole
symmetry of the theory at the quantum level.
We have to check that our renormalization procedure would be consistent
with the constraints which are the generators of the symmetry.
Consistency of the constraints requires
that commutators of the constraints do not
lead to new constraints.
The momentum constraints are generators of
three-dimensional general coordinate transformations.
Since our renormalization procedure preserves
three-dimensional general coordinate invariance, no anomalous terms may
appear in commutators  with the momentum constraints.
There remains to be considered only the commutator of
the Hamiltonian constraints.
The Hamiltonian constraint is given by
\be
{\cal H}(x) \equiv - \frac{16 \pi}{m_p^2}\, \Delta_{R}(x)
                   + \frac{m_p^2}{16\pi}\sqrt{h(x)}
                    \Bigl(\!~^{3}\!R(x) + 2 \Lambda \Bigr)\ .
\label{13}
\ee
More correctly, we compute the commutator
$ [ \int d^3x \xi(x) {\cal H}(x) , \int d^3x' \eta(x') {\cal H}(x') ]$
for arbitrary scalar functions $\xi$ and $\eta$.
An anomalous term could appear from the commutators of $\Delta_{R}$
and $\sqrt{h} ~^{(3)}R$.
In fact we find the following anomalous term:
\be
\phi^{(1)}(0) \int d^3x \sqrt{h}
   ( \xi \nabla\!_i\, \eta - \eta \nabla\!_i\, \xi ) \nabla^i\! ~^{3}\!R\ ,
\label{14}
\ee
with a nonzero coefficient.
This term may be expected from dimensional analysis
and the antisymmetry
under the exchange of $\xi$ and $\eta$.
We therefore take
\be
\phi^{(1)} (0) = 0\ ,
\label{15}
\ee
for consistency of the constraints with our renormalization procedure.

Now we want to solve the (renormalized) Wheeler-DeWitt equation (\ref{12})
with the condition (\ref{15}).
To solve it,
we attempt an expansion of the wave function of the universe
in powers of the Planck mass.
We assume that the wave function $\Psi[h]$ has the form
\be
\Psi[h] \equiv \exp \Bigl\{ -S[h]\Bigr\}
=  \exp \biggl\{ - \sum^{\infty}_{n=1}
       \biggl(\frac{m_p^2}{16\pi} \biggr)^{\!\!2n}
         \,S_n[h] \biggr\}\ \ .
\label{16}
\ee
Then, the Wheeler-DeWitt equation becomes
\be
\Delta_{R}(x)\, S
  - G_{ijkl}(x) \frac{\delta S}{\delta h_{kl}(x)}
    \frac{\delta S}{\delta h_{ij}(x)}
  + \biggl(\frac{m_p^2}{16\pi} \biggr)^{\!\!2} \sqrt{h(x)}
      \Bigl(\! ~^{3}\!R(x) + 2 \Lambda \Bigr) = 0\  ,
\label{17}
\ee
so that to leading order
\be
\Delta_{R}(x)\, S_1 = - \sqrt{h(x)}
         \Bigl(\! ~^{3}\!R(x) + 2 \Lambda \Bigr)\ .
\label{18}
\ee
We adopt an {\it ansatz} of locality\cite{Mans} to solve
the Wheeler-DeWitt equation in the expansion (\ref{16}).
We will assume that $S[h]$ is a sum of integrals of
local functions of $h_{ij}$.
Then, from eqs.(\ref{11}) with the condition (\ref{15}),
we find a solution to leading order as\footnote
{ In ref.\cite{Kodama},
Kodama has pointed out that the exponential of the Chern-Simons
action, which is equivalent to the Einstein action \cite{C-S gravity},
is an exact
solution of the Hamiltonian constraint in the holomorphic representation of
the Ashtekar formalism\cite{Ashtekar}.
Connections with our solution are unclear.}
\be
S_1 = \int d^3x \sqrt{h(x)} \Bigl\{ c_1 + c_2 ~^{3}\!R(x) \Bigr\}\ ,
\label{19}
\ee
where
\begin{eqnarray}
c_1 & = & - \frac{3!}{\al_1 \phi^{(3)}(0)}
          \biggl( 2 \Lambda - \frac{3! \bt_1 \phi^{(5)}(0)}
               {5! \bt_2 \phi^{(3)}(0)}\biggr)\ ,\nonumber\\
c_2 & = & - \frac{3!}{\bt_2 \phi^{(3)}(0)}\ .
\label{20}
\end{eqnarray}

Before we discuss physical interpretations of the wave function we
found to leading order,
we would like to explain
physical meanings of the expansion (\ref{16}).
The semiclassical expansion corresponds to the expansion
in powers of $\hbar$,
while the expansion (\ref{16}) corresponds
to the expansion in powers
of  $\hbar^{-2}$ because $\hbar$ appears in the combination of
$\hbar m_p^{-2}$
in quantum gravity.
Thus the wave function to leading order
in the expansion (\ref{16})
is expected to describe  quite different physics from the semiclassical one.
The semiclassical approximation
will be valid for long-wavelength gravitational fields,
while the expansion
(\ref{16}) may  be useful when the wavelength of gravitational fields
very rapidly changes in a wavelength because the first term
in eq.(\ref{17}) has to be more important compared with the second term.
The accuracy of our solution to leading order may be gauged
by comparing the
magnitudes of the successive terms
$(m_p^2/16 \pi)^{2} S_1$ and $(m_p^2/16 \pi)^{4} S_2$ in series for $S$.
Taking $\Lambda = 0$ for simplicity and assuming $\phi^{(n)}(0) \sim \mu^n$,
where $\mu$ is a mass parameter, we can write the leading term as
\be
\biggl(\frac{m_p^2}{16\pi}\biggr)^{\!\!2} S_1 = \mu^3 \int d^3x
\sqrt{h}
   \biggl(\frac{m_p}{\mu}\biggr)^{\!\!4}
   \biggl\{ a_1 + a_2 \frac{~^{3}\!R}{\mu^2} \biggr\}\ ,
\label{21}
\ee
where $a_1$ and $a_2$  are dimensionless constants of order one.
The next leading term may be of the form,
\begin{eqnarray}
\biggl(\frac{m_p^2}{16\pi}\biggr)^{\!\!4} S_2 &=&
   \mu^3 \int d^3x \sqrt{h}
     \biggl(\frac{m_p}{\mu}\biggr)^{\!\!8} \biggl\{  b_1
        + b_2 \frac{~^{3}\!R}{\mu^2}\nonumber\\
    & & + \frac{1}{\mu^4} \Bigl(  b_3(\! ~^{3}\!R)^2
            + b_4\!~^{3}\!R_{ij}\! ~^{3}\!R^{ij}
              + b_5 \nabla\!_i \nabla^i \!~^{3}\!R \Bigr) \biggr\}\  ,
\label{22}
\end{eqnarray}
where $b_n$'s are dimensionless constants.
It follows that we can drop the next leading term when $\mu \gg m_p $ and
$ ~^{3}\!R \sim \mu^2$, $\nabla\!_i \nabla^i \!~^{3}\!R \sim \mu^4$, etc.
We therefore expect that the wave function to leading  order
describes the universe with the curvature much larger than the
Planck scale.
If we apply the expansion (\ref{16}) to a minisuperspace model,
we can see that the expansion
is  essentially identical to a small radius expansion of the universe.
Thus we may expect that our expansion gives a good approximation
for the universe with the radius much smaller than the Planck scale,
though the minisuperspace model will not be applicable to this region.

Let us now discuss the physical interpretations of our result (\ref{19}).
We may compute expectation values of operators $\Omega$ for $\Psi$ as
\be
< \Omega >\ =\
\frac{\int{\cal D}h\, \Psi^{*}[h]\,\Omega\,\Psi[h]}
   {\int{\cal D}h\,\Psi^{*}[h]\,\Psi[h] }\ \ .
\label{23}
\ee
Since a precise definition of expectation values (or inner products) is not
known in quantum gravity, the right hand side of eq.(\ref{23})
is to be regarded
as a formal definition of expectation values.
If we put the wave function $\Psi[h]$ to leading order into eq.(\ref{23})
and note that $\Psi[h]$ is real,
we find that the expression is a path
integral representation for the vacuum expectation values
in three-dimensional Einstein gravity.
Therefore, we may conclude that
in the universe with the very small radius or with the very large curvature
beyond the Planck scale
expectation values
of operators can be calculated as the vacuum expectation values of
{\it three}-dimensional
quantum Einstein gravity.
Three-dimensional Einstein gravity has no local excitation, i.e.,
no graviton but a finite number of degrees of freedom,
and is in fact a topological field theory \cite{C-S gravity,Hosoya}.
An interesting speculation from the above observations is that
beyond the Planck scale
dimensional reduction
of four-dimensional quantum gravity occurs\footnote{Dimensional
reduction in quantum gravity has been discussed by 'tHooft
\cite{reduction} in a different context.}
and topological
phase appears \cite{topol}.
One might expect that the same mechanism of dimensional reduction could
occur for three-dimensional Einstein gravity.
However, this is not the case because of the topological nature of
the two-dimensional
Einstein-Hilbert action.

We would like to make comments on the factor-ordering problem
and the renormalizability.
We have chosen a factor-ordering shown in eq.(\ref{4}).
Other choices of factor-orderings will lead to different values
of numerical constants, e.g., $\al_n$'s and $\bt_n$'s in eq.(\ref{7})
but will not change the qualitative features of our results.
In our formulation, the renormalizability of the theory requires
that all physical quantities must be independent of the arbitrary function
$\phi$, so that the Newton \lq\lq constant" and the cosmological
\lq\lq constant" will be
functionals of $\phi$.
To leading order the above statement may be replaced by saying that
the wave function is independent of $\phi$.
This requirement implies that the Newton \lq\lq constant" $G=m_p^{-2}$
( the cosmological \lq\lq constant" $\Lambda$ ) is
a function of $\phi^{(3)}(0)$
($\phi^{(3)}(0)$ and $\phi^{(5)}(0)$).
We may define a $\bt$-function for $1/G^2$ as
\be
\bt \equiv \mu \frac{\partial}{\partial \mu}
      \Bigl(\frac{1}{G^2}\Bigr)\ ,
\ee
where $\mu \equiv ( \phi^{(3)}(0) )^{1/3}$ is a parameter of
mass dimension one.
Requiring that the wave function to leading order is independent of $\mu$
gives $\bt = 3/G^2$, which tells us that $1/{G(\mu)}^2$
is proportional to $\mu^3$.
The actual (dimensionless) expansion parameter is
$(m_p(\mu)/\mu)^4 = (G^2(\mu) \mu^4)^{-1}$,
which decreases as the mass scale $\mu$ increases.
Thus, the expansion (\ref{16}) is expected to give a good
approximation  scheme for large $\mu$.


\vspace{1cm}
\begin{center}
{\Large\bf Acknowledgements \par}
\end{center}

We would like to thank A. Hosoya, K. Inoue, M. Kato, Y. Kazama,
C. S. Lim
and T. Yoneya for useful discussions.


\end{document}